\pdfoutput=1
\documentclass[%
 reprint,superscriptaddress,
 amsmath,amssymb,
 aps,prb,floatfix,longbibliography
]{revtex4-2}
\usepackage{graphicx}
\usepackage{dcolumn}
\usepackage{bm}
\usepackage{multirow}

\usepackage{braket}
\usepackage{color}
\usepackage{ulem}
\usepackage{hyperref}

\usepackage{placeins}
\usepackage{bbm}
\usepackage{xcolor}

\newcommand{\calT}{\mathcal{T}}

\begin{document}


\title{Finite-size and finite bond dimension effects of tensor network renormalization}

\author{Atsushi Ueda}
\email{aueda@issp.u-tokyo.ac.jp}
\affiliation{%
 Institute for Solid State Physics, University of Tokyo, Kashiwa 277-8581, Japan
}%
\author{Masaki Oshikawa}
\affiliation{%
 Institute for Solid State Physics, University of Tokyo, Kashiwa 277-8581, Japan
}%
\affiliation{Kavli Institute for the Physics and Mathematics of the Universe (WPI),
The University of Tokyo, Kashiwa, Chiba 277-8583, Japan}
\affiliation{Trans-scale Quantum Science Institute, University of Tokyo, Bunkyo-ku, Tokyo 113-0033, Japan}

\date{\today}

\begin{abstract}
We propose a general procedure for extracting the running coupling constants of the underlying field theory of a given classical statistical model on a two-dimensional lattice,
combining tensor network renormalization (TNR) and the finite-size scaling theory of conformal field theory.
By tracking the coupling constants at each scale, we are able to visualize the renormalization group (RG) flow and demonstrate it with the classical Ising and 3-state Potts models.
Furthermore, utilizing the new methodology, we reveal the limitations due to finite bond dimension $D$ on TNR applied to critical systems.
We find that a finite correlation length is imposed by the finite bond dimension in TNR, and it can be attributed to an emergent relevant perturbation that respects the symmetries of the system.
The correlation length shows the same power-law dependence on $D$ as the ``finite entanglement scaling'' of the matrix product states.
\end{abstract}

\maketitle

\section{Introduction}

The universality of critical phenomena is one of the most intriguing and important concepts in statistical physics.
The renormalization group (RG), proposed by Wilson \cite{wilson1971renormalization,wilson1971renormalization2,wilson1975renormalization}, provides a conceptual framework to comprehend and characterize this universality.


In the RG framework, a universality class of critical phenomena is governed by an RG fixed point in the ``theory space''.
Theory space is the abstract space of all possible models or theories that can describe a physical system. Each point in this space represents a unique combination of the parameters of the theory, or more concretely, the corresponding Hamiltonian or action. In the context of the RG, we explore this “theory space" by starting from a specific point in the theory space and applying the RG transformations. These transformations effectively move us through the theory space, changing the values of the parameters as we coarse-grain the system. Importantly, models within the same universality class converge to an identical position through the RG transformations.
This allows for diverse critical phenomena to be comprehended in terms of perturbations to the fixed-point Hamiltonian and their respective scaling behavior.

This theoretical approach has directly facilitated the development of concrete schemes for the calculation of critical exponents. A prime example is the $\epsilon$-expansion for the $\phi^4$ theory in $4-\epsilon$ dimensions~\cite{RevModPhys.46.597,wilson1972critical}.
While the practical utility of such a scheme for calculating critical exponents may appear to be limited, it is imperative to underscore that the RG framework establishes the conceptual foundation for understanding the universality of critical phenomena.

In particular, the fixed point displays conformal invariance in two dimensions, thereby simplifying the associated theory which is described by a conformal field theory (CFT). The effective Hamiltonian near the RG fixed point Hamiltonian, denoted $\hat{H}_{\text{CFT}}$, can be expressed as follows:
\begin{align}
 \hat{H} &= \hat{H}_{\text{CFT}} + \sum_j g_j \int_0^L dx \; \hat{\Phi}_j(x),
\label{eq:CFT_perturbed}
\end{align}
In this expression, $\hat{\Phi}_j(x)$ represents a scaling operator with a scaling dimension $x_j$, and $g_j$ denotes the corresponding coupling constant. In two dimensions, the running coupling constants $g_j$ are renormalized as $g_j \propto L^{2-x_j}$ as functions of a scale $l=\ln(L/a)$.
In general, there are only a few \textit{RG-relevant} coupling constants with $x_j <2$, which increase as the scale $l$ increases.

There also exists \textit{RG-irrelevant} coupling constants with $x_j>2$ that decrease as the scale $l$ increases. Despite their occasional significance, the principal characteristics of critical phenomena can be outlined primarily by considering the limited number of RG-relevant coupling constants. Differential equations, termed RG equations, frequently describe the evolution of these running coupling constants as functions of the scale $l$. Field theory methods frequently serve as the basis for deriving these RG equations.

However, it is worth noting that the exact determination of RG equations may not always be feasible when the corresponding field theory is not solvable. Moreover, the application of RG to lattice models has generally been challenging for quantitative calculations. While offering an intuitive understanding of RG, the “block spin transformation" method falls short as a practical computational method for lattice models. Overall, early RG schemes for lattice models saw limited success, the notable exception being Wilson's numerical renormalization group for impurity problem~\cite{wilson1975renormalization}.
 
Subsequently, density matrix renormalization group(DMRG)~\cite{refs_on_DMRG} emerged as a highly effective numerical algorithm for one-dimensional quantum many-body systems. Despite its name, DMRG is typically employed as a numerical algorithm with less emphasis on RG flows in the ``theory space.''

More recently, the development of tensor network renormalization(TNR)~\cite{PhysRevLett.99.120601,PhysRevLett.115.180405,PhysRevB.95.045117,PhysRevLett.118.110504,PhysRevLett.118.250602,PhysRevB.97.045111} opened a way to implement numerical schemes for a wide range of lattice models in a manner more faithful to the original concept of RG. Notably, it is possible to obtain a fixed-point tensor after multiple iterations of TNR steps. This fixed-point tensor encapsulates critical information about the infrared (IR) fixed point, including conformal data.

Regarding RG flow, there have been numerous previous studies\cite{PhysRevLett.121.230402,PhysRevLett.115.180405,PhysRevLett.118.110504,PhysRevLett.118.250602,Delcamp_2020}.
Yet, a generic and quantitative framework for calculating RG flows remains elusive, primarily due to challenges in maintaining the correlation between the changes in numerically obtained tensor networks and the RG flow within the 'theory space.'

\par{} In this paper, we first propose an efficient and quantitative scheme to extract the RG flow numerically from TNR, discussed in Sec.~\ref{sec:RGflow}. This involves comparing the finite-size spectrum of the transfer matrix with CFT.
Concrete examples, such as the numerical results of the Ising and 3-state Potts models, are employed to validate the theoretical predictions. Our method also provides an efficient and accurate estimation of the critical point, extending the 'Level Spectroscopy' technique previously developed for Berezinskii-Kosterlitz-Thouless (BKT) transitions~\cite{nomura1994critical,PhysRevB.104.165132}.

\par{} 
Leveraging this methodology, we uncover the effects of finite bond dimension $D$ on TNR at criticality in Section \ref{section_fixed_point}.
The finite-bond approximation of tensors constrains the effective correlation length, preventing the attainment of a 'true fixed point tensor' corresponding to a nontrivial RG fixed point through repeated TNR procedures. 
While this phenomenon was reported in earlier studies \cite{PhysRevLett.99.120601,PhysRevLett.115.180405,PhysRevB.95.045117,PhysRevLett.118.110504,PhysRevLett.118.250602,PhysRevB.97.045111}, it has been often overlooked.
Our numerical results suggest that the finite bond-dimension effects can be attributed to an emergent relevant perturbation that respects the symmetry of the lattice model. 
Furthermore, we demonstrate that the finite correlation length that is imposed by the finite bond dimension scales in the same way as in matrix product states (MPS).

We note that some of the methods and observations discussed in this paper were previously reported in our earlier publication~\cite{PhysRevB.104.165132}, where they were applied to the classical XY model.
The goal of the present paper is to illustrate the more widespread applicability of this approach and deliver a more comprehensive analysis of the finite bond-dimension effects.\\

Sample codes necessary to reproduce the figures presented in this paper, along with introductory reviews on TRG and TNR, are accessible via Jupyter notebooks at the following GitHub repository: \url{https://github.com/dartsushi/Loop-TNR_RGflow}.

\section{Review on tensor network renormalization and conformal field theory}
\begin{figure}[tb]
    \centering
    \includegraphics[width=86mm]{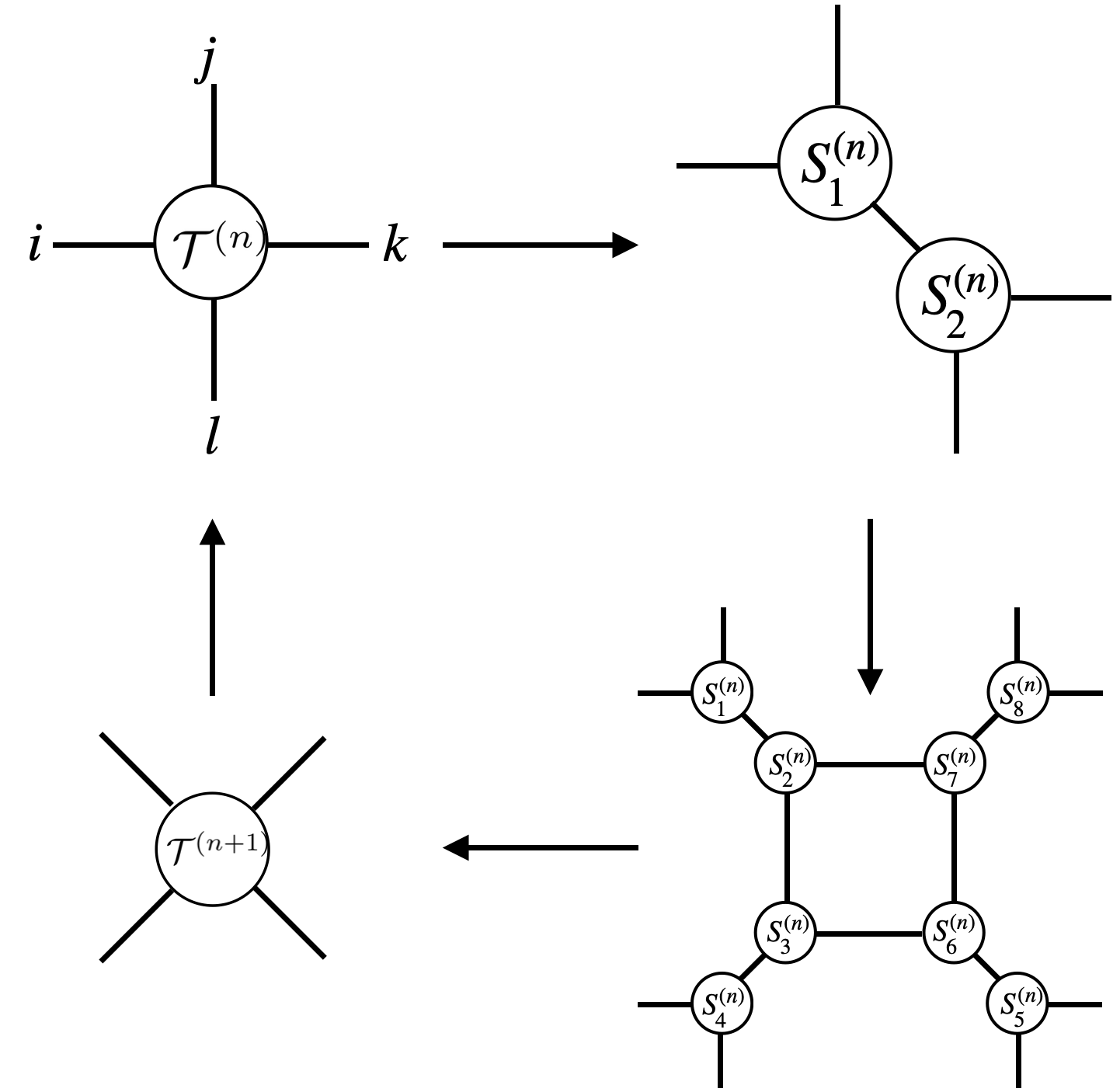}
    \caption{A schematic picture of the tensor network renormalization. The effective local Boltzmann weight at $n$-th RG step $\calT^{(n)}$ is decomposed into the two three-leg tensors and recombined as $\calT^{(n+1)}$. The effective system size enlarges by $\sqrt{2}$ each RG step. The typical bond dimension and the maximum number of RG steps employed in this paper are $D\leq 40$, and RG steps $\leq 30$, respectively.}
    \label{tnr}
\end{figure}
\subsection{Tensor network renormalization}
The tensor network is a numerical technique used to represent the partition function of statistical models. The partition functions of two-dimensional statistical models with a system size of $L$ can be expressed through the contraction of $L^2$ tensors. Each tensor represents a local Boltzmann weight, and its dimensions correspond to physical degrees of freedom. For instance, the local tensor of the Ising model on the square lattice is a four-leg tensor $\calT^{(1)}_{ijkl}=e^{\beta(s_is_j+s_js_k+s_ks_l+s_ls_i)}$. The tensor network representation often provides an efficient method for simulating complex systems.\\
However, the exact contraction of $L^2$ tensors is generally impracticable for larger system sizes due to the constraints imposed by the high-dimensional Hilbert space. TNR aims to circumvent this issue by utilizing the principles of renormalization group theory. During each step of the RG process, $\calT^{(n)}$ is coarse-grained to $\calT^{(n+1)}$ via a series of decompositions and recombinations, as illustrated in Fig.~\ref{tnr}.
Starting from the local tensor $\calT^{(1)}$, we can simulate a system size of $L=\sqrt{2}^n$ after $n$ RG steps. The process of coarse-graining in TNR involves numerical truncation, reducing the number of degrees of freedom while preserving essential physics. Consequently, TNR facilitates efficient numerical simulation of complex systems.

\subsection{Critical phenomena under conformal invariance}
The following sections mainly discuss the Ising and 3-state Potts models on the square lattice. The energy (classical Hamiltonian) of the Ising and 3-state Potts models are 
\begin{align}
    \mathcal{E}_{\rm Ising} &= -\sum_{\langle i,j\rangle}s_is_j-h\sum_is_i,\\
    \mathcal{E}_{\rm Potts} &= -\sum_{\langle i,j\rangle}\delta_{s_i,s_j},
\end{align}
where $s_i=\pm 1$(Ising) and $s_i=0,1,2$(3-state Potts). The first terms and $h$ represent the nearest-neighbor interactions and the magnetic field.
Employing the temperature $T$, the Boltzmann weight is defined as $e^{-\mathcal{E}/T}$, where we set the Boltzmann constant to unity.
Our primary focus in the main text is the Ising model, while a detailed discussion of the 3-state Potts model is provided in the appendix.
The Ising model reaches its critical point at $(T,h)=(T_c,0)$, where $T_c=2/\ln{(1+\sqrt{2})}$.
At this criticality, physical quantities like the spin-spin correlation function are governed by the Ising CFT, which comprises three primary operators: the identity operator $I$, magnetic operator $\sigma$, and energy operator $\epsilon$.

In the context of the lattice model, a shift from the critical temperature and the application of a magnetic field correspond to the perturbative insertion of $\epsilon$ and $\sigma$ into the effective Hamiltonian. As a result, $\sigma$ is odd in the $\mathbb{Z}_2$ spin-flip, while $I$ and $\epsilon$ are even. Given the operator structure of the CFT, certain quantities are consequently fixed.

The two-point correlation function is defined as
$$\langle\Phi_i(r_i)\Phi_j(r_j)\rangle = \frac{\delta_{i,j}}{|r_i-r_j|^{2x_i}},$$
where $\Phi_i$ represents a primary operator, and $x_I = 0$, $x_\sigma = \frac{1}{8}$, and $x_\epsilon = 1$ are the scaling dimensions. In a similar vein, the three-point correlation function adopts a universal form, represented as
$$\langle\Phi_i(r_i)\Phi_j(r_j)\Phi_k(r_k)\rangle = \frac{C_{ijk}}{|r_i-r_j|^{\Delta_{ij}^k}|r_j-r_k|^{\Delta_{jk}^i}|r_k-r_i|^{\Delta_{ki}^j}},$$
where $C_{ijk}$ is an operator product expansion(OPE) coefficient, and  $\Delta_{ij}^k = x_i+x_j-x_k$. 
The non-zero OPE coefficients are given as follows:
\begin{align}
    C_{III}&=C_{I\sigma\sigma} = C_{I\epsilon\epsilon} =1,\\
    C_{\sigma\sigma\epsilon}&=\frac{1}{2}.
\end{align}
The permutation of the indices does not change the OPE coefficients\footnote{The combination of the indices in non-zero $C_{ijk}$ preserves $\mathbb{Z}_2$ symmetry.}.(For further details of CFT, we suggest readers see Ref.~\cite{francesco2012conformal}.) The collection of information on the scaling dimension $x_i$ and $C_{ijk}$, referred to as the CFT data, is crucial to understanding critical phenomena. As such, determining the CFT from a numerical standpoint is of paramount importance.

\section{Computation of field-theory data and RG flow from TNR}
\label{sec:RGflow}

\subsection{Scaling dimensions}
\label{subsec:scaling_dim}

For simplicity, let us consider a classical statistical model on the square lattice with nearest-neighbor interactions only.
Then the local Boltzmann weight can be represented by a tensor $\calT$ with four open indices, and the partition function is given by
contraction of a tensor network which consists of the tensor $\calT$.

More specifically, the partition function $Z(L_x,L_y)$ for the system of the size $L_x \times L_y$ is given by the contraction of
the network of $L_x \times L_y$ identical tensors $\calT$.

Under a single step of TNR, the length scale represented by a single tensor is renormalized by $\sqrt{2}$. After $N$ steps, the renormalized tensor becomes $\calT(L)$, which represents the length scale $L = \sqrt{2}^N$.
As $\calT(L)$ is equivalent to the $L\times L$ contracted tensor network up to truncation errors, contractions of the horizontal and vertical {legs} yield the partition function $Z(L,L)$ in periodic boundary condition (PBC). Similarly, contracting only the legs in the $x$-direction gives the $L$-stacks of the transfer matrix in $y$-direction. Since one can regard the transfer matrix as the imaginary-time evolution operator of corresponding one-dimensional quantum systems, its eigenvalues $\lambda_n(L)$ are related to the energy levels $E_n(L)$ of the quantum system as $\lambda_n(L) = \exp\left(-LE_n(L)\right)$. For convenience, we define the rescaled energy levels $x_n(L)$ by $E_n(L)-E_0(L) =2\pi x_n(L) /L$\footnote{In the classical systems, the {characteristic} velocity $v$, playing the role of the speed of light, is unity because the system invariant under the exchange of the $x$ and $y$ axes.}
to obtain
\begin{align}
\frac{\lambda_n(L)}{\lambda_0(L)}=\exp(-2\pi x_n(L))\label{TNR_spectrum}.
\end{align}
Exactly at the criticality, this rescaled energy level $x_n(L)$ coincides with the scaling dimension $x_n$ of the in the thermodynamic limit $(L\rightarrow\infty)$~\cite{cardy1984conformal,cardy1986operator,PhysRevB.80.155131}.

If the system is off-critical and without a spontaneous symmetry breaking,
the rescaled energy level of the ``first excitation'' is asymptotically proportional to the system size as $\Delta L/(2\pi)$, where
$\Delta$ is the inverse correlation length (excitation gap) in the thermodynamic limit.

Summarizing these observations, naively speaking, we can judge whether the system is critical or not by
looking at the asymptotic behavior of the rescaled energy levels $x_n(L)$.
If they grow linearly in $L$, the system is off-critical.
If they approach to constants, the system is critical, and the scaling dimensions of the operators can be read off from the
asymptotic values of $x_n(L)$ in the thermodynamic limit.
While this can be a useful guide, there are corrections from RG-irrelevant perturbations, and more importantly, due to the limitation of a finite bond dimension,
as we will discuss later.

\subsection{Operator product expansion coefficients}
Operator product expansion is another fundamental concept in field theory and statistical mechanics~\cite{kadanoff,wilsonOPE}. Since OPE coefficients determine the structure of the field theory, their computation is quite important. Numerical computation of OPE coefficients~\cite{PhysRevLett.116.040401,PhysRevResearch.4.023159} has not been so straightforward compared to that of
scaling dimensions. Here, we present a simpler way to compute them, which is applicable to TRG~\cite{PhysRevLett.99.120601}, HOTRG~\cite{PhysRevB.86.045139}, and Loop-TNR~\cite{PhysRevLett.118.110504}. \\

As explained in the previous section, the renormalized tensor $\calT^{(N)}$ contracted in $x$-direction is a transfer matrix in the $y$-direction. While the eigenvalues of the transfer matrix correspond to the energy or scaling dimension of the primary operators, the eigenvectors thereof are the wavefunctions of the corresponding
``primary states'' $|\psi_n(L)\rangle$. This is graphically represented below.
\begin{align*}
    \includegraphics[width=60mm]{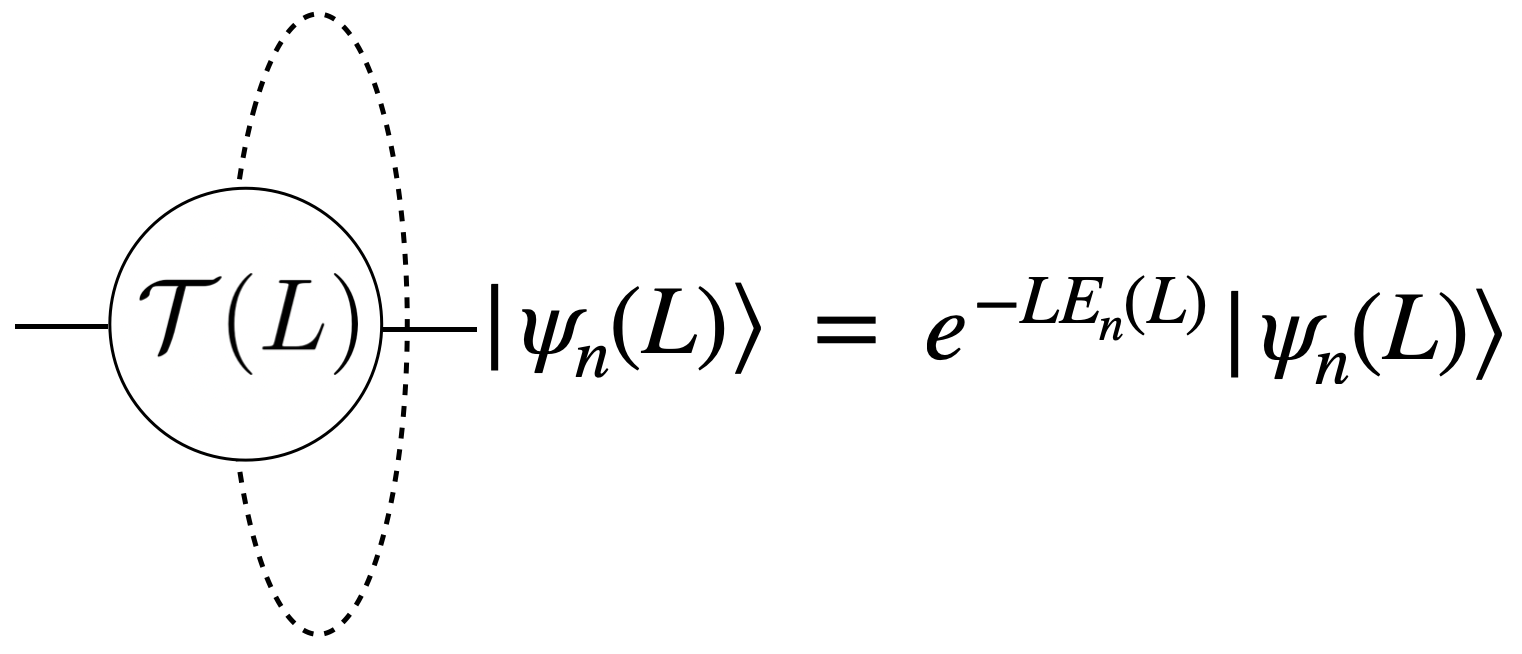}
\end{align*}
Note that the tensor has been rotated for ease of viewing. We do not change the contracted index. Likewise, we can compute the wavefunctions of the system size $2L$ as depicted below. 
\begin{align*}
    \includegraphics[width=60mm]{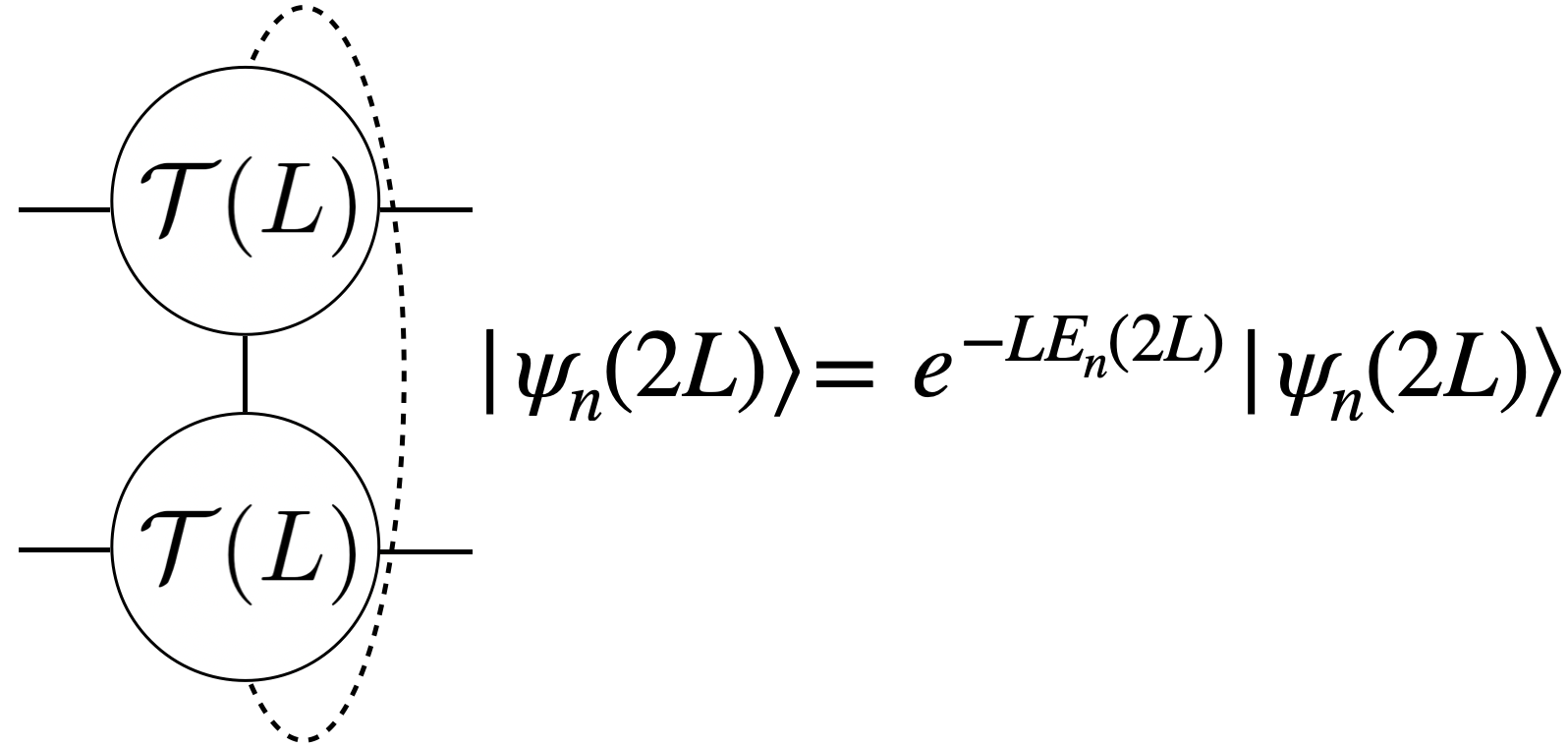}
\end{align*}
$|\psi_n(L)\rangle$ and $|\psi_n(2L)\rangle$ are one-leg and two-leg tensors, respectively. Thus, one can calculate the overlaps $\ket{\psi_\alpha(2L)}$ and $\ket{\psi_\beta(L)}\otimes\ket{\psi_\gamma(L)}$ by contracting the indices. 
\begin{table}[tb]
\begin{ruledtabular}
\begin{tabular}{lllll}
\multicolumn{1}{c}{$\psi_\alpha$} & \multicolumn{2}{c}{$\psi_\beta,\psi_\gamma$} & \multicolumn{1}{c}{$C_{\alpha\beta\gamma}$} & \multicolumn{1}{c}{$2^{2\Delta_\beta+2\Delta_\gamma-\Delta_\alpha}{A_{\alpha\beta\gamma}}/{A_{III}}$} \\\hline
$I$                                 & \multicolumn{2}{l}{$\sigma,\sigma$}          & 1                                           & 0.8938                                                     \\
$\sigma$                          & \multicolumn{2}{l}{$\sigma,I$}               & 1                                           & 0.9473                                                     \\
$I$                                 & \multicolumn{2}{l}{$\epsilon,\epsilon$}      & 1                                           & 0.9966                                     \\
$\epsilon$                        & \multicolumn{2}{l}{$\epsilon,I$}             & 1                                           & 0.9968                                                     \\
$\epsilon$                        & \multicolumn{2}{l}{$\sigma,\sigma$}          & 0.5                                         & 0.5007                                                      \\
$\sigma$                          & \multicolumn{2}{l}{$\sigma,\epsilon$}        & 0.5                                         & 0.2705                                                     
\end{tabular}
\end{ruledtabular}
\caption{The numerically obtained OPE coefficients of the Ising CFT from TRG. The bond dimension and the system size are $D=56$ and $L=16\sqrt{2}$(9 RG steps), respectively.\label{OPE_table} }
\end{table}

In CFT, the overlap $\langle{\psi_\alpha(2L)}|{\psi_\beta(L)}{\psi_\gamma(L)}\rangle$ is proportional to the “pants diagram" of path integrals~\cite{PhysRevB.105.125125,PhysRevB.105.165420,arxiv.2203.14992}, and the OPE coefficient and the overlap are related as
\begin{align}
    \frac{A_{\alpha\beta\gamma}}{A_{III}} =
    2^{\Delta_\alpha-2\Delta_\beta-2\Delta_\gamma}C_{\alpha\beta\gamma},\label{OPE_eq}
\end{align}
where $A_{\alpha\beta\gamma}$, $\Delta,$ and $I$ are $\langle{\psi_\alpha(2L)}|{\psi_\beta(L)}{\psi_\gamma(L)}\rangle$, the scaling dimension, and the identity operator, respectively. In most cases, the identity operator corresponds to the ground state.
We benchmark our method by the critical Ising model. Table.~\ref{OPE_table} shows the numerically obtained OPE coefficients by TRG~\cite{PhysRevLett.99.120601} at $L=16\sqrt{2}$ and $D=56$.
Naturally, there are finite-size corrections to Eq.~\eqref{OPE_eq}.
Since Eq.~\eqref{OPE_eq} is exact in the thermodynamic limit, using a very large system size $L$ might appear desirable.
However, as we will discuss later in Sec.~\ref{section_fixed_point}, corrections due to the finite bond-dimension effect appear for system sizes larger than a correlation length $\xi(D)$\footnote{This effect is even stronger and non-trivial for TRG}. 
As reported in Ref.~\cite{arxiv.2203.14992}, the finite-size effects are significant for $C_{\sigma\sigma\epsilon}$ and $C_{\epsilon\epsilon I}$.
Nevertheless, even with the moderate size $L=16\sqrt{2}$, the obtained values $C_{I\epsilon\epsilon} = 0.9966$ and $C_{\epsilon\sigma\sigma}=0.5007$ are rather close to exact CFT results. While we tested our method by the simplest algorithm, Levin and Nave's TRG, the method for calculating OPE is straightforwardly applicable to other TRG and TNR algorithms, such as HOTRG~\cite{PhysRevB.86.045139}.

\subsection{Level Spectroscopy}
\label{sec:LevelSpectroscopy}

As we have mentioned earlier, the rescaled energy levels $x_n(L)$ in Eq.~\eqref{TNR_spectrum} would be
independent of the scale $L$ and give the scaling dimensions of the corresponding operators, if the system
were exactly described by a CFT.
However, the rescaled energy levels of a lattice model generally depend on the system size $L$,
as the effective Hamiltonian of the system contains perturbations to the CFT as in Eq.~\eqref{eq:CFT_perturbed}.

The rescaled energy levels in a finite-size perturbed CFT are given as~\cite{cardy1984conformal,cardy1986operator} 
\begin{align}
x_n(L)=x_n+2\pi\sum_jC_{nnj}g_j(L),\label{fss_scaling_dimension}
\end{align}
where $g_j(L)$ scales as $\propto L^{2-x_j}$. 
Comparing Eq.~\eqref{TNR_spectrum} from TNR and Eq.~\eqref{fss_scaling_dimension} from the conformal perturbation theory,
we can obtain the running coupling constants $g_j(L)$ at each scale from the finite-size effect $\delta x_n(L)=x_n(L)-x_n$.

An immediate application of this observation is an accurate determination of the critical point.
While such a framework is dubbed ``level spectroscopy''
was developed for BKT transition, which is notoriously difficult for
standard finite-size scaling analysis, first for quantum spin systems in one dimension~\cite{nomura1994critical}
and recently extended for classical statistical systems in two dimensions using TNR~\cite{PhysRevB.104.165132},
the basic idea is also applicable to more conventional critical phenomena such as in the Ising model.

The RG fixed point for the two-dimensional Ising model has two relevant operators, the energy density $\epsilon$
and the magnetization density $\sigma$.
The coupling constant $g_\epsilon$ for $\epsilon$ is proportional to the deviation of the temperature from the critical point,
and also scaled $\sim L$ in the small coupling limit $g_\epsilon \ll 1$ because $x_\epsilon=1$.
Thus
\begin{align}
g_{\epsilon}(L) \sim \alpha (T-T_{c}) L ,
\label{eq:linear_app}
\end{align}
when $g_\epsilon(L) \ll 1$.
Likewise, the coupling $g_\sigma$ is proportional to the magnetic field $h$ and scaled $\sim L^{15/8}$ because $x_\sigma=1/8$.

Although the Ising critical phenomena are mostly described by the two relevant coupling constants $g_\epsilon$ and $g_\sigma$,
more accurate description can be obtained by including irrelevant perturbations.
Including the leading irrelevant operators, namely the irrelevant operators with the smallest scaling dimension permitted by the symmetries,
we obtain
\begin{align}
    H=H^{*}_{\rm Ising}+\int_0^Ldx&[g_\sigma\sigma(x)+g_\epsilon\epsilon(x)\nonumber\\
    &+g_{T^2} T_{\rm cyl}^2(x)+g_{\bar{T}^2}\bar{T}_{\rm cyl}^2(x)],
\end{align}
where $T_{\text{cyl}}$ and $\bar{T}_{\text{cyl}}$ are the holomorphic and anti-holomorphic parts of stress tensor on a cylinder~\cite{cardy1986operator}.
The holomorphic part $T_{\text{cyl}}$ of the stress tensor on a cylinder is related to that on the infinite plane
$T_{zz}(z)$ via the conformal mapping $z = e^{2\pi w/L}$, where $w= \tau + ix$ and $0 \leq x < L$. More explicitly, $T_{zz}(z)$ transforms as
\begin{align}
    T_{\rm cyl}(w) &= \left( \frac{2\pi}{L}\right)^2 \left( z^2 T_{zz} (z)-\frac{c}{24} \right) .
\end{align}
This leads to
\begin{align}
    T_{\text{cyl}}(x) &= \frac{2\pi}{L} \left( \sum_{n=-\infty}^\infty L_n e^{2\pi i x/L} - \frac{c}{24} \right),
\end{align}
where $c$ is the central charge characterizing the CFT, and
$L_n$'s are generators of the Virasoro algebra defined by
\begin{align}
    T_{zz}(z) = \sum_{n=-\infty}^\infty \frac{L_n}{z^{n+2}},
\end{align}
in terms of the holomorphic part $T_{zz}$ of the energy-momentum tensor on the infinite plane.
Inserting the above $T_{\text{cyl}}$ and integrating over $0 \leq x < L$ with an appropriate regularization,
the $g_{T^2}$-term of the perturbation is given as~\cite{poghosyan2019shaping}
\begin{align}
   \int dx \; T^2_{\rm cyl}(x) =L_0^2-\frac{c+2}{12}L_0+2\sum_{n=1}^{\infty}L_{-n}L_{n}+\frac{c(22+5c)}{2880}
   \nonumber
\end{align}
Only the first and second terms affect the energy levels, and the contributions to $x_\sigma(L)$ and $x_\epsilon(L)$ are calculated to be
$-\frac{7}{768}g_{T^2}$ and $\frac{7}{48}g_{T^2}$ respectively.
The computation of the contributions from $\bar{T}^2$ is exactly the same, and we denote their sum as $g$.
These operators are the leading irrelevant operators for the Ising model on the square lattice.
Although they break the continuous rotation symmetry (which corresponds to the Lorentz
invariance in the Minkowski space-time), they are allowed on the square lattice, which is invariant only under the discrete C$_4$ rotation.
The calculation of $x_\sigma(L)$ and $x_\epsilon(L)$ is straightforward, and they are shown in Table.~\ref{RGscalingdimension}~\footnote{As $T_{\text{cyl}}^2$ and $\bar{T}_{\text{cyl}}^2$ are not primary operators, we need to pay special attention. The details are discussed in the Appendix.}. 

While the exact critical point is known for the Ising model on the square lattice, let us demonstrate
the determination of the critical point from the TNR spectrum without using prior knowledge of the critical point
(but utilizing the CFT data, assuming that we identify the universality class).
Since we are interested in the critical point at zero magnetic fields, we can set $g_\sigma \propto h =0$.
The simplest way to determine the critical point is to look at the lowest rescaled energy level
$x_\sigma(L)$ in the lowest order of the relevant coupling constant $g_\epsilon$,
ignoring the irrelevant perturbation $g$.
Within this approximation, the shift $\delta x_\sigma(L) = x_\sigma(L) - x_\sigma$ vanishes at the critical point $T=T_c$ where $g_\epsilon = 0$.
Away from the critical point, $\delta x_\sigma(L)$ is non-zero and grows proportionally to $L$ because $g_\epsilon(L)$ scales as $L$.
Because of this, we can identify the critical point with the temperature where $\delta x_\sigma(L)=0$ is observed in the TNR spectrum.
However, this estimate suffers from the corrections due to the leading irrelevant perturbations $T_{\text{cyl}}^2$ and $\bar{T}_{\text{cyl}}^2$.
Since they have scaling dimension $4$, the corresponding coupling constant is renormalized as $g \propto L^{-2}$.
This leads to an error of $O(L^{-2})$ in the naive estimate of the critical point using $\delta x_\sigma(L)=0$.

\begin{figure}
    \centering
    \includegraphics[width=86mm]{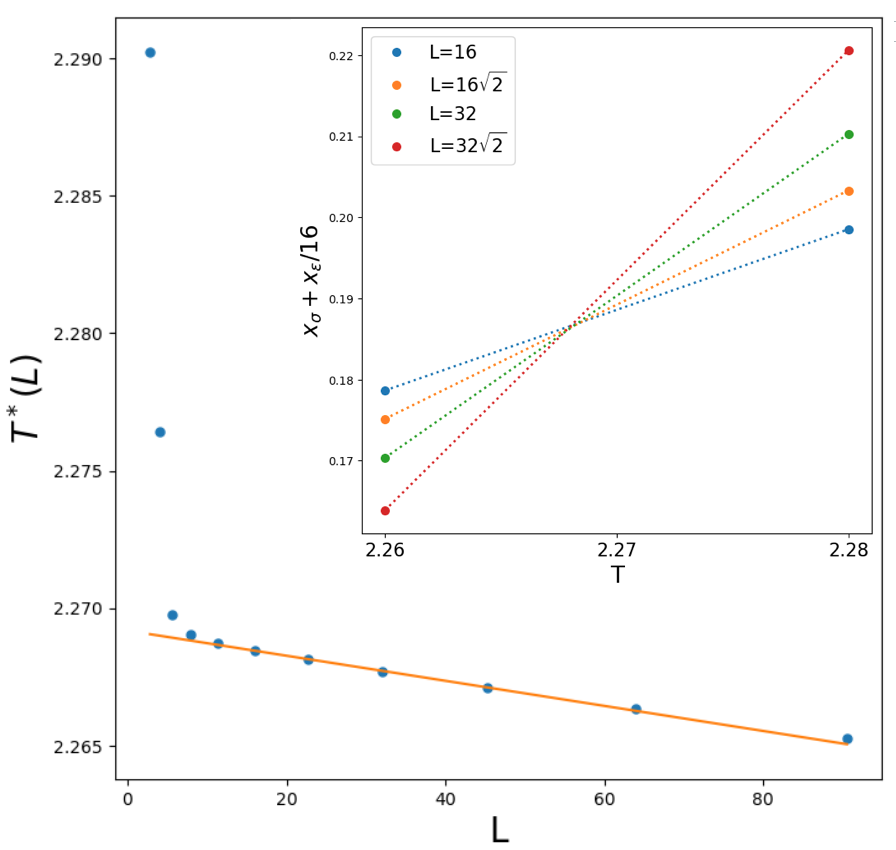}
    \caption{Example of estimating the transition temperature using Loop-TNR. We set $T^-=2.66$ and $T^+=2.68$ as an initial estimate. The level-crossing temperature $T^*(L)$ is linearly fitted to extrapolate the transition temperature. The insert shows how we compute $T^*(L)$ for various system sizes.}
    \label{Tc_extrapolation}
\end{figure}

We can improve the accuracy by removing the effects of the leading irrelevant perturbation $g$.
This can be done by combining the shifts of the rescaled energy levels $\delta x_\sigma(L)$ and $\delta x_\epsilon(L)$ as
\begin{align}
\delta x_\text{cmb} \equiv & \delta x_\sigma(L)+\frac{1}{16}\delta x_\epsilon(L)\nonumber\\
    &=
    \pi g_\epsilon + 
    (\alpha^\sigma_\sigma+\frac{1}{16}\alpha_\epsilon^\sigma) g_\sigma^2 +(\alpha^\epsilon_\sigma + \frac{1}{16}\alpha^\epsilon_\epsilon ) g_\epsilon^2.
\label{eq:deltax_combined}
\end{align}
Note that the first-order correction in the irrelevant coupling $g$ is canceled out.
Now we can identify the critical point by finding the temperature for which $\delta x_\text{cmb} \propto g_\epsilon(L) = 0$.
Having eliminated the effects of the leading irrelevant perturbation $T_{\text{cyl}}^2, \bar{T}_{\text{cyl}}^2$, the dominant error is now caused by the next-leading irrelevant
operator with scaling dimension $6$ and thus should be scaled as $L^{-4}$.

In practice, the determination of the critical point can be efficiently implemented as follows.
First, we pick up one temperature from each phase: $T^{+} > T_{c}$ and $T^{-} < T_{c}$, and calculate the combined shift
$\delta x_\text{cmb}$ at these temperatures. 
The phase of the system can be confirmed by observing the growth of $\delta x_\text{cmb}$ as the system size increases because it increases/decreases if the system is in the high-temperature/low-temperature phase (if the initial choice of the temperature turns out to be wrong, change the temperature and restart the process).
Next, linear interpolations of the combined shift between the two temperatures $T^{\pm}$ are made,
and the crossing of the lines for system sizes $L$ and $\sqrt{2}L$ is found, as shown in the insert of Fig.~\ref{Tc_extrapolation}.
We denote the temperature where the two lines cross as $T^{*}(L)$.
Because of the second-order contribution $O({g_\epsilon}^2)$ in Eq.~\eqref{eq:deltax_combined},
the crossing temperature $T^*(L)$ obtained by the \textit{linear} interpolation deviates from the true critical point $T_c$
as $T^{*}(L) - T_{c} \propto g_\epsilon \propto L$, when $g_{\epsilon} \ll 1$\footnote{It is proportional to $L^{2-x_{\text{thermal}}}$, where $x_{\text{thermal}}$ is the scaling dimension of the thermal operator.}.
The critical point $T_c$ is estimated by fitting $T^{*}(L)$ by a linear function of $L$ as $T^*(L) \sim T_c + \text{const.} L$.
While the ``extrapolation'' to $L=0$ used here might look unusual, this procedure is done to remove the effect of the nonlinearity due to $O({g_\epsilon}^2)$
in Eq.~\eqref{eq:deltax_combined}, and the condition $\delta x_\text{cmb} = 0$ itself is accurate for $T_c$ up to the error of $O(L^{-4})$ due to the next-leading irrelevant
perturbations.
An example of the estimate of $T_c$ with the above procedure with the choice of the temperatures $T^{+}=2.68$ and $T^{-}=2.66$
and with system sizes $16 \leq L < 64$ is depicted in Fig.~\ref{Tc_extrapolation}.
The final estimate of the critical point is $T_c^\text{est} = 2.269177$.
Remarkably, even with the choice of two temperatures differ by $10^{-2}$ and the relatively low bond-dimension $D=20$, the estimated critical point is quite accurate:
$T_c^\text{est}- T_{c} = -8.11 \times 10^{-6}$.
This is thanks to the suppression of the error to $O(L^{-4})$ by eliminating the contributions from the leading irrelevant operators.
Once the critical point is estimated with good accuracy with this procedure, the accuracy can be further improved
by choosing $T^{\pm}$ closer to the estimated critical temperature and then applying the same procedure.

\begin{table}[tb]
\begin{ruledtabular}
\begin{tabular}{c|c|l|c}

           model                    & \multicolumn{2}{c|}{operator}             & Rescaled energy level                \\ \hline
\multirow{3}{*}{Ising model}
& \multicolumn{2}{c|}{ } & \\[0.5ex]
& \multicolumn{2}{c|}{
$x_\sigma(L)$ }&
$\frac{1}{8}+\alpha_\sigma^\sigma g_\sigma^2+\pi g_\epsilon+\alpha^\epsilon_\sigma g_\epsilon^2-\frac{7}{768}\pi g$\\[1.5ex]
&\multicolumn{2}{c|}{$x_\epsilon(L)$}&$1+\alpha^\sigma_\epsilon g_\sigma^2+\alpha^\epsilon_\epsilon g_\epsilon^2+\frac{7}{48}\pi g$\\[1.5ex] 
\end{tabular}
\end{ruledtabular}
\caption{The finite-size scaling dimension of the Ising model. $\alpha$ is a constant determined from the second-order perturbation. Since $g_{T^2}$ and $g_{\bar{T}^2}$ decay in the same manner, we write them as $g$.\label{RGscalingdimension}}
\end{table}

\subsection{Renormalization Group flow}

\begin{figure*}[tb]
    \centering
    \includegraphics[width=178mm]{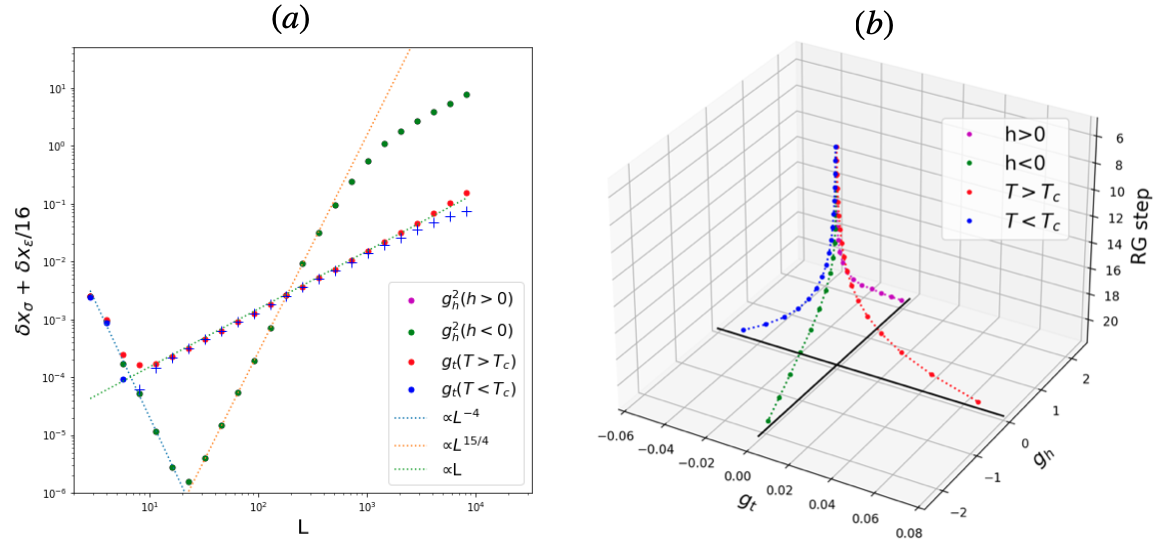}
    \caption{(Left panel) The system size dependence of $\delta x_\text{cmb} = \delta x_\sigma + \delta x_\epsilon/16$ for $h=\pm10^{-5}$(purple and green), $T= 1.0001T_c$(red) and $T= 0.9999T_c$(blue). The purple and green dots are on top of each other, and “$+$" denotes the data with a negative sign. After removing the $L^{-2}$ irrelevant perturbations, the next leading $L^{-4}$ perturbation shown with a blue dotted line appears. The data was obtained via Loop-TNR with a bond dimension of $D = 24$, which was deemed sufficient for the finitely-correlated systems being considered. (Right panel) The resulting renormalization group flow. Only data after six steps are exhibited, where the $L^{-4}$ perturbations disappear.}
    \label{Flow_Ising}
\end{figure*}

The comparison between the TNR spectrum~\eqref{TNR_spectrum} and the conformal perturbation theory~\eqref{fss_scaling_dimension}
can also be used to extract running coupling constants and their scale dependence, enabling a visualization of the RG flow.
While this was shown for the BKT transition in the XY model~\cite{PhysRevB.104.165132},
here let us demonstrate the method for the Ising model.
This will also be useful to investigate the finite bond-dimension effects in detail, as we will discuss in Sec.~\ref{section_fixed_point}.
The extraction of running coupling constants in the Ising model is again based on the shifts of the rescaled energy levels in Table~\ref{RGscalingdimension},
and it is useful to consider the combined shift~\eqref{eq:deltax_combined} also for this purpose.
Given $g_\sigma$ and $g_\epsilon$ are small in the vicinity of criticality, we neglect $g_\epsilon^2$ for $h=0$ and redefine two relevant coupling constants as $g_t=\pi g_\epsilon$ and $g_h=\sqrt{(\alpha^\sigma_\sigma+\frac{1}{16}\alpha_\epsilon^\sigma)} g_\sigma$ for convenience.
In this way, the combined shift Eq.~\eqref{eq:deltax_combined} simply gives $g_t$ when $h=0$ and ${g_h}^2$ when $T=T_c$, in the lowest order of $g_t, g_h$.
Using these relations, we can read off the relevant coupling constants $g_t$ or $g_h$ from the TNR data, as shown in Fig.~\ref{Flow_Ising}($b$).
As we have discussed in the previous subsection, the effects of the leading irrelevant perturbations $T_{\text{cyl}}^2, \bar{T}_{\text{cyl}}^2$ with scaling dimension $4$
are eliminated in the combined shift~\eqref{eq:deltax_combined}, and thus the finite-size correction is now of $O(L^{-4})$, due to the next-leading
irrelevant operators with scaling dimension $6$.
This $O(L^{-4})$ scaling is indeed observed in Fig.~\ref{Flow_Ising} near the critical point for small system size $L$ when relevant perturbations are still negligible.
Since it is safe to say that these contributions disappear after five RG steps, we can conclude that the origin of $g_t$ and $g_h$ are purely from $\epsilon$ and $\sigma$ after six steps. 
\par{} The right panel illustrates the scale-dependence of the coupling constants $g_t$ and $g_h$. It is nothing but the RG flow of the Ising critical point, and we conclude that we succeed in calculating the RG flow of the celebrated Ising fixed point.
\par{}There is one thing to note on the left panel of Fig.~\ref{Flow_Ising}.
While the combined shift~\eqref{eq:deltax_combined}, which is an estimator for $|g_h|^2$,
scales as $L^{3.75}$ at $L<10^3$, it starts to flatten and scales as $L$ at $L>10^3$.
This behavior has a rather simple origin.
Since the magnetic perturbation is relevant, the system has a finite correlation length or equivalently, a non-zero gap $\Delta$.
This implies that the rescaled energy levels are proportional to $L$ for sufficiently large system size $L \gg \Delta^{-1}$,
as mentioned in Sec.~\ref{subsec:scaling_dim}.
As a consequence, the shift~\eqref{eq:deltax_combined} also grows proportionally to $L$.
In this regime, the conformal perturbation theory breaks down (higher-order contributions are important), and we no longer
identify the shift~\eqref{eq:deltax_combined} with $|g_h|^2$.
This should be distinguished from the $L$-linear behavior of the combined shift~\eqref{eq:deltax_combined}
observed for $L>10$ with $h=0$ and $T \neq T_c$, which corresponds to the renormalization of $g_t \propto L$ because of $x_\epsilon=1$.
The $L$-linear behavior due to the gap is observed in the non-perturbative regime $\delta x_{\epsilon, \sigma} \gg x_{\epsilon,\sigma}$, whereas
the $L$-linear behavior due to the scaling is observed in the perturbative regime $\delta x_{\epsilon, \sigma} \ll x_{\epsilon,\sigma}$.

\section{Finite bond-dimension effects\label{section_fixed_point}}

Let us examine the impacts of a finite bond-dimension $D$ on TNR from the perspective of our method.
In any computation that employs tensor networks, it is necessary to restrict the bond dimension to a finite value $D$ due to the increasing storage requirements and computational costs associated with larger bond dimensions.
The finiteness of the bond dimension inevitably leads to a loss of information in each step of renormalization after a certain number of iterations.
Although TNR can nominally handle arbitrary large systems, and the TNR-type calculations are often used to study extremely large systems,
we have to be careful about the limitations due to the finite bond dimension.

The limitation of the finite bond dimension $D$ on the MPS is characterized by the finite (maximum) correlation length $\xi(D)$
of the MPS~\cite{luca_fes,pollmann2009theory,Pirvu_fes}.
The correlation length of MPS is known to obey the scaling law
\begin{align}
    \xi(D) \sim & D^\kappa ,
    \label{eq:xi_D_scaling}
    \\
    \kappa = & \frac{6}{c(1+\sqrt{\frac{12}{c}})} .
    \label{eq:kappa_c}
\end{align}
While the TNR-type calculation of two-dimensional statistical systems appears rather different from the MPS applied to one-dimensional quantum systems,
the emergence of the finite correlation length $\xi(D)$ obeying the similar scaling law~\eqref{eq:xi_D_scaling} was reported in
Ref.~\cite{PhysRevB.89.075116} for a HOTRG calculation of the critical Ising model in two dimensions.
The exponent $\kappa$ for the Ising model was estimated to be approximately $2$, which is close to the MPS exponent~\eqref{eq:kappa_c} $\kappa = 2.03425\ldots$
for the Ising CFT with central charge $c=1/2$.
A similar emergence of the finite correlation length $\xi(D)$ was also reported in our TNR finite-size scaling study of the two-dimensional XY model~\cite{PhysRevB.104.165132},
with the MPS exponent~\eqref{eq:kappa_c} for $c=1$.

In the following, using our TNR finite-size scaling methodology, we will demonstrate that the emergence of the finite correlation
length due to the finite bond dimension in TNR can be attributed to an emergent relevant perturbation (Sec.~\ref{sec:emergent_perturbation}).
Furthermore, we present evidences for the scaling~\eqref{eq:xi_D_scaling} with the MPS exponent~\eqref{eq:kappa_c}
in TNR of Ising and 3-state Potts models (Sec.~\ref{sec:scaling_xi}).

\subsection{Emergent relevant perturbation}
\label{sec:emergent_perturbation}

\begin{figure}[tb]
    \centering
    \includegraphics[width=86mm]{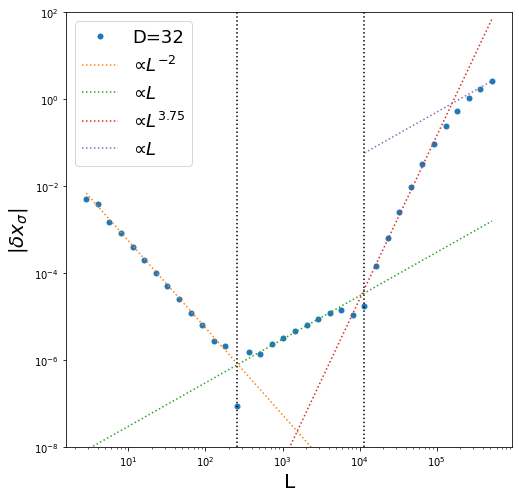}
    \caption{Shift $|\delta x_\sigma(L)|$ for the Ising model at $T=T_c$, $h =0$ computed by Loop-TNR with $D=32$.
    There is little finite-$D$ effect for small system sizes $L<256$. The emergent perturbations of $\epsilon$ and $\sigma$ appear at $L\sim 256$ and $L\sim10^4$, scaling as $L$ and $L^{15/4}$. The induced gap by finite-$D$ goes towards constant at $L>10^5$ as denoted with the purple dotted line.}
    \label{mixed_emergent_perturbation}
\end{figure}

If a finite correlation length emerges in the TNR, it would be
natural to identify the renormalized tensor with a Hamiltonian for the system away from the critical point,
that is, an RG fixed-point (CFT) Hamiltonian perturbed with relevant operators
\begin{align}
H_{\text{FB}}(D)=H^*_{\text{CFT}}+\sum_i\int_0^L dxg_i(D,L)\Phi_{i}(x,D),
\end{align}
where $H_{FB}$ is the effective Hamiltonian of the finite-$D$ system and $\Phi_{i}(x,D)$ are the scaling operators representing the perturbations.
In this view, we expect relevant perturbations to emerge in order to mimic the finite correlation length imposed by the finite bond dimension.

To demonstrate the emergence of the relevant perturbation, we investigate the system-size dependence of the
shift in the rescaled energy levels $\delta x_\sigma$.
In Fig.~\ref{mixed_emergent_perturbation}, we show the absolute value of the shift $|\delta x_\sigma|$ as a function of the system size $L$
used in calculating the transfer matrix spectrum in TNR exactly at the critical point $h=0, T=T_c$.
The conformal perturbation theory in Eq.~\eqref{fss_scaling_dimension} implies that
the shift $x_\sigma$ contains contributions from the irrelevant perturbations.
Since the leading irrelevant operators at the critical points are $T_{\text{cyl}}^2$ and $\bar{T}_{\text{cyl}}^2$ with scaling dimension $4$,
we expect $\delta x_\sigma(L)$ decays as $L^{-2}$. 
(This is to be contrasted with Eq.~\eqref{eq:deltax_combined} and Fig.~\ref{Flow_Ising}, in which the contributions from $T_{\text{cyl}}^2$ and $\bar{T}_{\text{cyl}}^2$
are eliminated.)
The expected $L^{-2}$ behavior in the shift $\delta x_\sigma(L)$ is indeed
observed for small system sizes $L<256$.
For larger system sizes, however, $|\delta x_\sigma(L)|$ starts to 
increase, deviating from the conformal perturbation theory scaling $L^{-2}$.
We identify the finite bond-dimension $D$ effects as the
origin of this deviation.
More remarkably, we can observe a clear scaling behavior of the deviation.
That is, the shift $|\delta x_\sigma(L)|$
 scales with the system sizes as $L$ and $L^{15/4}$ for $256<L<10^4$ and $10^4<L$, respectively.
Compared with the off-critical cases in Fig.~\ref{Flow_Ising}, we realize that these scalings are identical to
those induced by the thermal and magnetic perturbations.
In other words, the relevant perturbations emerge in the TNR calculation.

Let us first discuss the $L^{15/4}$ scaling of the shift, observed for $L > 10^4$.
This can be understood as the effect of an emerging magnetic perturbation $h$.
Although the magnetic perturbation $h$ is forbidden by the $\mathbb{Z}_2$ spin-flip symmetry,
the symmetry could be broken by the limitations in the machine precision.
Once the spin-flip symmetry is broken,
the magnetic field $h$, which is a relevant perturbation, is effectively generated.
Even if the effective magnetic field $h$ is extremely small, it will be enhanced at each RG step and
eventually dominates the system at sufficiently large length scales.
This is what we observe for $L > 10^4$.
This phenomenon should be related to machine precision and not intrinsic to the algorithm.
If we are interested in a $\mathbb{Z}_2$ symmetric system, we can impose the symmetry at each step of TNR
in order to avoid this effect.

In contrast, the $L$ scaling observed for $256 < L < 10^4$ is more intrinsic.
The most relevant perturbation allowed under the $\mathbb{Z}_2$ symmetry to the critical Ising fixed point
is the thermal operator.
Thus, we expect that the finite bond dimension effect can be mimicked by the thermal perturbation $\epsilon$
to the fixed-point Hamiltonian $H^*_\text{CFT}$.
If this is the case, the effective coefficient $g_\epsilon$ grows proportionally to $L$ as the system size $L$
is increased, because the thermal operator $\epsilon$ has the scaling dimension $1$.
According to Eq.~\eqref{fss_scaling_dimension}, this will lead to a correction proportional to $L$
in the rescaled energy level $\delta x_\sigma(L)$.
This is indeed supported by the numerical result shown in Fig.~\ref{mixed_emergent_perturbation}.

In general, the finite-$D$ effect in TNR would be described in terms of the emergence of relevant perturbation(s) to the fixed-point Hamiltonian,
which induces the finite correlation length $\xi(D)$.
In addition to the emergence of the relevant operator $\epsilon$ in the critical Ising model discussed above, a similar emergence of the relevant operator is observed in the critical 3-state Potts model, as demonstrated in Appendix~\ref{Potts_sec}.

\subsection{Scaling of the emergent correlation length}
\label{sec:scaling_xi}

\begin{figure*}[tb]
\begin{center}
\includegraphics[width=178mm]{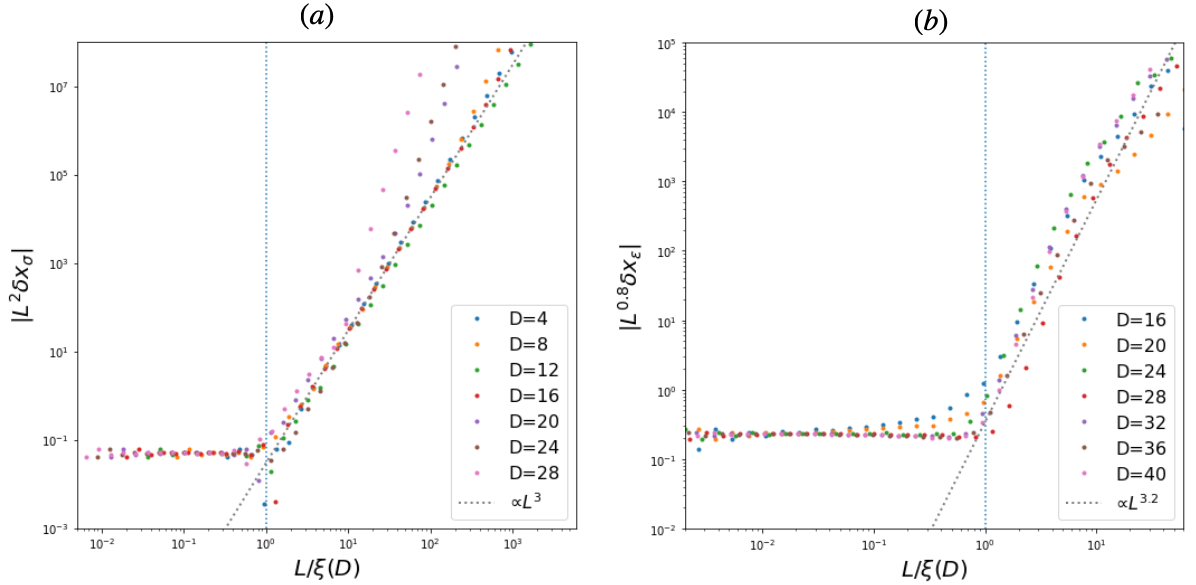}
 \caption{
 $(a)$ The scaling of the shift $\delta x_\sigma$ in TNR of the Ising model at the critical point, for various bond dimensions $D=4, \ldots, 28$.
 The vertical axis is scaled as $L^2 \delta x_\sigma$ so that it is constant when $L \ll \xi(D)$.
 When $L \ll \xi(D)$, the shift is dominated by the emergent relevant perturbation $\epsilon$; this is confirmed by the scaling $L^2 g_\epsilon \propto L^3$.
 The horizontal axis is scaled as $L/\xi(D)$, where the correlation length $\xi(D)$ is hypothesized as in Eqs.~\eqref{eq:xi_D_scaling} and~\eqref{eq:kappa_c}.
 The collapse of the data for different bond dimensions is strong evidence of the hypothesized scaling of the correlation length $\xi(D)$.
The blue dotted line indicates $L/\xi(D)=1$. We set $\xi(D) =2.0 D^\kappa$ so that $L/\xi(D)=1$ becomes the crossover scale between the finite-size scaling regime and the finite-$D$ scaling regime.
 $(b)$ Similar scaling analysis of the shift $\delta x_\epsilon$ in TNR of the 3-state Potts model at the critical point, for various bond dimensions $D=16, \ldots, 40$ with $\xi(D) =0.067 D^\kappa$.
 The scaled shift $L^{0.8} \delta x_\epsilon$ behaves as a constant in the finite-size scaling regime $L/\xi(D) <1$, whereas it scales as $L^{3.2}$ in the finite-$D$ scaling regime
 $L/\xi(D) > 1$, as expected from the CFT analysis (see Appendix~\ref{Potts_sec} for details).
 The data for different bond dimensions collapse again, giving compelling evidence for the scaling of the correlation length~\eqref{eq:xi_D_scaling} and~\eqref{eq:kappa_c}
 }
 \label{FES_Ising}
 \end{center}
\end{figure*}

Now let us demonstrate that the finite correlation length $\xi(D)$ induced by the finite bond dimension $D$ in TNR
obeys the same scaling~\eqref{eq:xi_D_scaling} and~\eqref{eq:kappa_c} as in the MPS, as suggested in
Refs.~\cite{PhysRevB.89.075116,PhysRevB.104.165132}.

In Fig.~\ref{FES_Ising}, we demonstrate the scaling of the correlation length induced by the finite bond dimension in TNR of the critical Ising and the 3-state Potts models.
In Fig.~\ref{FES_Ising}(a), we plot the shift $\delta x_\sigma$ in the Ising model obtained by the TNR of the Ising model at the critical point, which was also studied in Fig.~\ref{mixed_emergent_perturbation}, with the several different bond dimensions $D=4, \ldots, 28$.
Here, we rescaled the vertical axis as $L^2 \delta x_\sigma$ so that the constant behavior is observed for system size smaller than the correlation length, where the leading irrelevant perturbation (which causes $\delta x_\sigma \propto L^{-2}$) is dominant.
The deviation from the constant at larger system sizes $L$ can be attributed to the emergent relevant perturbation
$\epsilon$ induced by the finite bond dimension $D$, as discussed in the previous subsection.
This is confirmed by the $L^3$ scaling ($L^2$ times $\delta x_\sigma \propto g_\epsilon \propto L$).
Most importantly, the horizontal axis is the rescaled system size $L/\xi(D)$ using the hypothesized correlation length $\xi(D)=aD^\kappa$ given by Eqs.~\eqref{eq:xi_D_scaling} and~\eqref{eq:kappa_c}.
The collapse of the data for different bond dimensions strongly supports our hypothesis on the correlation length. Note that we roughly fit the prefactor $a$ so that the cross-over occurs at $L=\xi(D).$ 

In order to confirm the finite-$D$ scaling of the correlation length and its universality, we have also studied the 3-state Potts model
at the critical point.
As an example, in Fig.~\ref{FES_Ising}(b), we plot the shift of the rescaled energy level corresponding to the energy operator $\epsilon$ in the 3-state Potts model.
For this shift $\delta x_\epsilon$, the contribution from the leading irrelevant operator is $\sim L^{-4/5}$, and the dominant contribution from the emergent
relevant perturbation $\epsilon$ is expected to be proportional to ${g_\epsilon}^2 \propto L^{12/5}$. (See Appendix~\ref{Potts_sec} for details).
We rescaled the vertical axis as $L^{0.8} \delta_\epsilon$ so that it is constant in the finite-size scaling regime $L < \xi(D)$.
The horizontal axis is again the rescaled system size $L/\xi(D)$, with the correlation length $\xi(D)$ defined in Eqs.~\eqref{eq:xi_D_scaling} and~\eqref{eq:kappa_c} with
the central charge $c=4/5$ for the 3-state Potts model.
The data for different bond dimensions again show a collapse, providing compelling evidence for our hypothesis on the correlation length scaling.
For $L/\xi(D) > 1$, the data fits well the expected behavior $L^{0.8} \times {g_\epsilon}^2 \propto L^{0.8} \times L^{2.4} = L^{3.2}$.

\section{Conclusion and discussion}
In the first part of the paper, we discussed a method for computing the coupling constants using renormalized tensors based on the finite-size scaling theory of CFT. By plotting the resulting values at each scale, we were able to visualize the RG flow, and we confirmed that the theoretical RG flows, as shown in Fig.~\ref{Flow_Ising}, are consistent with the Ising and XY models. Our method has the advantage of being able to extract both ultraviolet and infrared information, making it a valuable tool for investigating gapped and crossover systems.
\par{} 
In the second part of the paper, applying the methodology developed in the first part, we explored the impact of finite bond-dimension $D$ on the RG flow.
The finiteness of the bond-dimension results in a finite correlation length $\xi(D)$,
or equivalently in a non-zero gap in the energy spectrum of the corresponding one-dimensional quantum system.
We find that this gap formation can be attributed to the emergence of a relevant perturbation enforced by the finite bond dimension.
This is demonstrated by the RG flow of the emergent relevant coupling.

The finite-size scaling of TNR shows a crossover at $L \sim \xi(D)$, above which the system is governed by the finite correlation length.
The correlation length $\xi(D)$ induced by the finite bond dimension in TNR
shows the same scaling~\eqref{eq:xi_D_scaling},~\eqref{eq:kappa_c} as the correlation length of MPS.
While such scaling in TNR was suggested earlier in Refs.~\cite{PhysRevB.89.075116,PhysRevB.104.165132}, in this paper, we presented more convincing evidence.

\begin{figure}[tb]
    \centering
    \includegraphics[width=86mm]{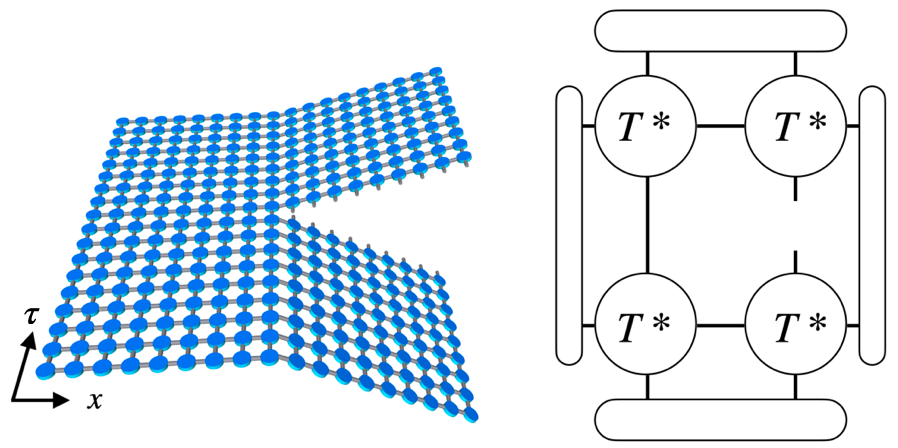}
    \caption{
    (Left panel) A schematic picture of the reduced density matrix $\rho_A$ for a bipartition of the system in the path integral picture.
    The uncontracted legs correspond to the indices of the reduced density matrix.
  (Right panel) Each of the four quadrants of the space-time in the left panel may be replaced by the renormalized tensor in TNR with appropriate boundary conditions.  }
    \label{reduced_rho}
\end{figure}

Although we do not have a mathematical proof for the scaling of $\xi(D)$ in TNR at this point, it may be natural from the following point of view.
Besides the construction of the transfer matrix by contracting horizontal legs, the renormalized tensor obtained in TNR can give the
corner transfer matrix by contracting the upper and left legs.
The same finite-$D$ scaling~\eqref{eq:xi_D_scaling},~\eqref{eq:kappa_c} as in MPS was observed in corner transfer matrix renormalization
group (CTMRG)~\cite{NISHINO199669,PhysRevE.96.062112,PhysRevE.101.062111}.
Moreover, the entanglement spectrum for the half-bipartition of the system of length $2L$ can be related to
a contraction of four renormalized tensors of linear size $L$~\cite{CalabreseLefevre2008}, as shown in Fig.~\ref{reduced_rho}.
These relations are suggestive of the identical scaling of $\xi(D)$ in MPS, CTMRG, and TNR as we have observed.

Our study highlights the importance of considering the impact of the finite bond dimension in the TNR-type approach.
In particular, a direct study of the thermodynamic limit with TNR would be prone to errors due to the finite correlation length $\xi(D)$ imposed by the finite bond dimension.
As a resolution of this problem, we have demonstrated that accurate data for the thermodynamic limit can be extracted by finite-size scaling of TNR spectra obtained for
system sizes smaller than $\xi(D)$, combined with conformal field theory. 
Even with this limitation, the tractable system size is greatly increased from $\sim \log{D}$ with exact diagonalization to $\xi(D) \sim D^\kappa$ in TNR.\\

\FloatBarrier

\textit{Note added}
When this work was almost completed, a closely related work~\cite{PhysRevB.107.205123} based on a HOTRG study of the Ising model appeared.
It is quite similar in spirit to this work, combining finite-size scaling and HOTRG.
Their estimate of the correlation length $\xi$ from the transfer matrix eigenvalue, and the determination of the critical point
based on the finite-size scaling of $\xi$, are essentially equivalent to our analysis of $\delta x_\sigma$ discussed in Sec.~\ref{sec:LevelSpectroscopy}.
Utilizing the knowledge of Ising CFT, we have further improved the accuracy by analyzing $\delta x_\text{cmb}$
which removes the effects of the leading irrelevant operators.
The error in the estimated transition temperature $T_c^\text{est}$ we obtained in Sec.~\ref{sec:LevelSpectroscopy} is about $10^{-5}$,
which is larger than theirs ($10^{-7} \sim 10^{-6}$).
However, our estimate is based on the data at two temperatures $T^\pm$ separated by $10^{-2}$ and can be further improved by taking more data points.
The MPS scaling~\eqref{eq:xi_D_scaling} and~\eqref{eq:kappa_c} of the correlation length in TNR we have discussed is also supported by them.

\section*{Acknowledgements}
A. Ueda thanks Tsuyoshi Okubo and Luca Tagliacozzo for stimulating discussions.
This work was supported in part by MEXT/JSPS KAKENHI Grant Nos. JP17H06462 and JP19H01808, and JST CREST Grant No. JPMJCR19T2,
and was partially done during
the program ``Tensor Networks: Mathematical Structures and Novel Algorithms'' held in
Erwin Schrödinger International Institute for Mathematics and Physics (ESI) at the University of Vienna.
A part of the computation in this work has been done
using the facilities of the Supercomputer Center, the Institute for Solid State Physics, the University of Tokyo.

\bibliography{references}
\appendix

\begin{table}[tb]
\begin{ruledtabular}
\begin{tabular}{ccc}
Symbol       & Dimension      & Meaning     \\ \hline
$I$ & 0              & identity    \\
$\epsilon$   & $\frac{2}{5}$  & thermal op. \\
$\sigma$     & $\frac{1}{15}$ & spin        \\
$X$          & $\frac{7}{5}$  &             \\
$Y$          & ${3}$  &             \\
$Z$          & $\frac{2}{3}$  &             \\ 
\end{tabular}
\caption{A set of primary operators of the 3-state Potts model.\label{potts_operators}}
\end{ruledtabular}
\end{table}
\section{Finite-Entanglement scaling of the Three-State Potts model.\label{Potts_sec}}
\begin{figure*}[tb]
    \centering
    \includegraphics[width=178mm]{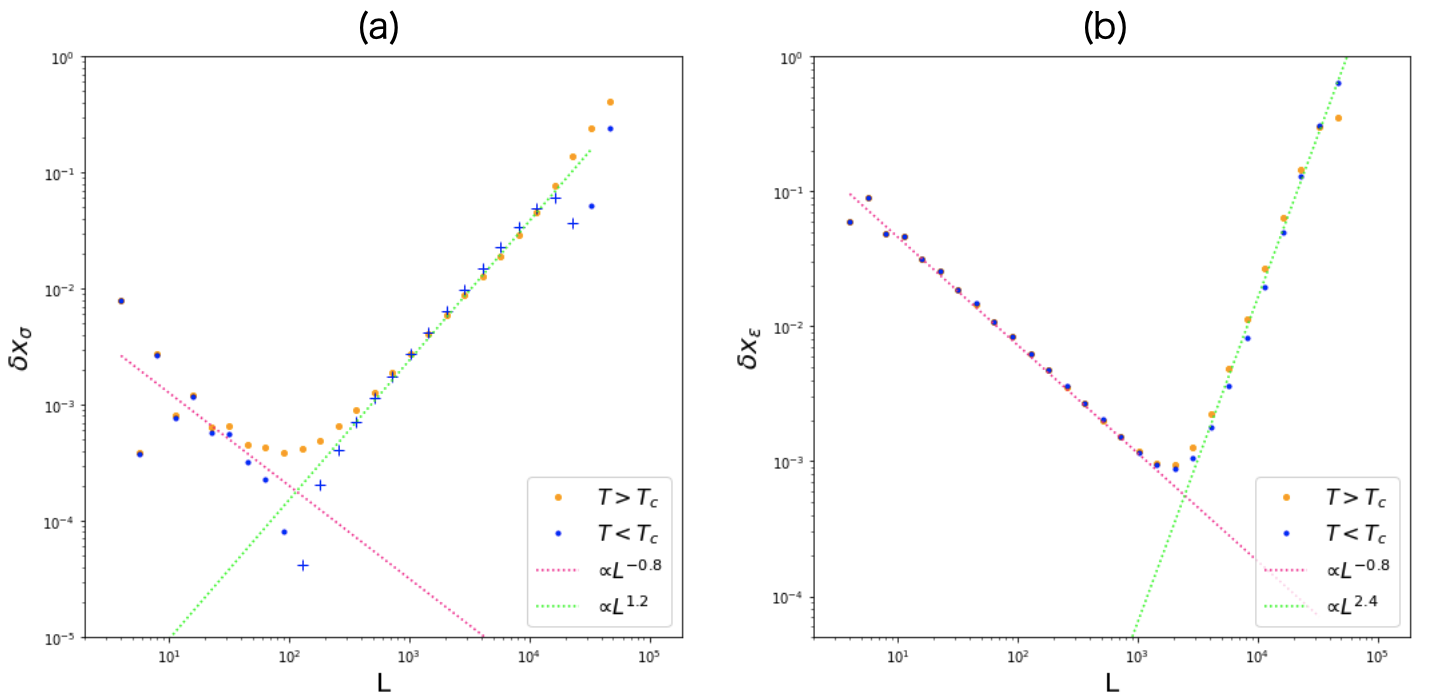}
    \caption{The size dependence of the (a)$\delta{x}_\sigma$ and (b)$\delta{x}_\epsilon$ at $T=0.999995T_c$ and $T=1.000005T_c$. The pink and green dotted lines denote $L^{-0.8}$, (a)$L^{1.2}$, and (b)$L^{2.4}$ fittings respectively. For the low-temperature phase, the sign of $\delta{x}_\sigma$ is negative at $L>100$. The dip on the left panel around $L\sim10^2$ corresponds to the zero point of Eq.~\eqref{Potts_perturbation}. (b) The finite-size effect to the $x_\epsilon$ suffers less from
    $T_{\text{cyl}}^2+\bar{T}_{\text{cyl}}^2$ in amplitude. The scaling of Eq.~\eqref{Potts_perturbation2} is clearly observed.}
    \label{Potts_sigma}
\end{figure*}
\subsection{Model}
We can further verify the emergence of relevant perturbations by applying it to the three-state Potts model. It is a natural extension of the Ising model to the $\mathbb{Z}_3$ symmetry, and the Hamiltonian is 
\begin{align}
    H=-\sum_{\langle i,j\rangle}\delta_{s_i,s_j},
\end{align}
where $s_i$ takes 0, 1, and $-1$. It has a phase transition of $\mathbb{Z}_3$ symmetry breaking at $T_c=1/\log(1+\sqrt{3})$.
The critical theory of the 3-state Potts model is another type of the minimal model $\mathcal{M}(6,5)$ with $c=\frac{4}{5}$\citep{francesco2012conformal,DOTSENKO198454}. A set of primary operators are shown in Table.~\ref{potts_operators}. 

As opposed to the Ising model, there are off-diagonal operators as $\Phi_{\frac{2}{5},\frac{7}{5}},\ \Phi_{\frac{7}{5},\frac{2}{5}}$ and $\Phi_{3,0},\ \Phi_{0,3}$(currents). 
\par{} Let us first examine the RG flow in a gapped system. Similar to the Ising model, the phase transition is identified by spontaneous symmetry breaking. The high-temperature phase is a trivial phase, whereas the low-temperature region is $\mathbb{Z}_3$ symmetry breaking phase. Thus, the fixed-point tensor is a stacking of three states with their $\mathbb{Z}_3$ charge $0,\ -1$, and 1.
\subsection{Construction of the effective Hamiltonian}
\par{}The RG flow can be seen by investigating the scaling dimensions. For instance, we can take the spin operator $\sigma=\Phi_{\frac{1}{15},\frac{1}{15}}$ and plot the value of $\delta{x}_\sigma=x_\sigma(L)-\frac{2}{15}$. Similarly, as in the Ising model, there is competition between irrelevant and relevant operators: $X=\Phi_{\frac{7}{5},\frac{7}{5}}$ and $\epsilon=\Phi_{\frac{2}{5},\frac{2}{5}}$. The thermal operator separates the $\mathbb{Z}_3$ symmetry-breaking phase from the trivial one. The finite-size corrections of $X$ and $\epsilon$ to $x_\sigma$ are $L^{-0.8}$ and $L^{1.2}$, respectively.
The fusion rules are $\sigma\times\sigma=1+\epsilon+\sigma+ X + Y + Z $, $\epsilon\times\epsilon=1+X$, and $\epsilon\times\sigma=\sigma+Z$.
Hence, $\delta{x}_\sigma$ has the following form:
\begin{align}
    \delta{x}_\sigma=2\pi c_{\sigma\sigma X} g_X\left(\frac{L}{2\pi}\right)^{-0.8}+2\pi c_{\sigma\sigma \epsilon} g_\epsilon\left(\frac{L}{2\pi}\right)^{1.2}. \label{Potts_perturbation}
\end{align}
On the other hand, the perturbation of $\epsilon$ appears as a second-order term for $\delta x_\epsilon$ because the fusion rule says $\epsilon\times\epsilon=1+X$. Consequently, $\delta x_\epsilon$ can be computed as
\begin{align}
\delta{x}_\epsilon=2\pi c_{\epsilon\epsilon X} g_X\left(\frac{L}{2\pi}\right)^{-0.8}+\alpha g^2_\epsilon \left(\frac{L}{2\pi}\right)^{2.4},\label{Potts_perturbation2}
\end{align}
where $\alpha$ is a constant determined from the second-order calculation.
\begin{figure}[tb]
    \centering
    \includegraphics[width=86mm]{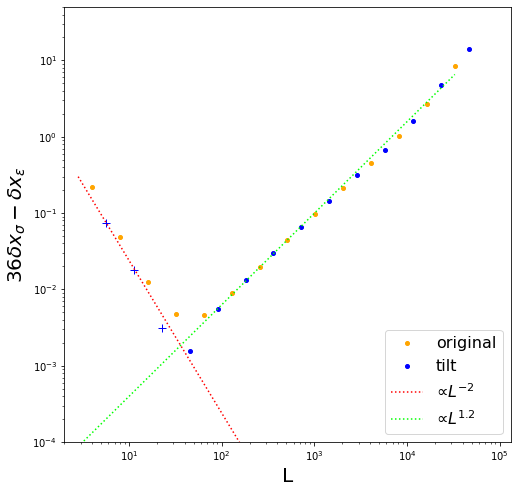}
    \caption{$36\delta x_\sigma-\delta x_\epsilon$ for the high temperature phase. “+” is used when the sign is negative. The red dotted line denotes the $L^{-2}$ fitting while the light green one is just a relevant $L^{1.2}$ contribution from $\epsilon$. Loop-TNR rotates the lattice by $\frac{\pi}{4}$ at each RG step, and the tilted system is plotted with the blue dots\label{Potts_irrelevant}.}
\end{figure}

Figure.~\ref{Potts_sigma} shows the computed $\delta x_\sigma$ by TNR. As expected, it exhibits the competition between irrelevant and relevant operators. The sign of $g_\epsilon$ is the opposite between two phases, which is a manifest indication of the RG flow in the opposite direction due to the thermal operator. $x_\sigma$ has doubly degenerate states with $\mathbb{Z}_3$ charge $\pm 1$. In the low-temperature phase, these two states flow to $x_\sigma(L)\rightarrow 0$, and the fixed point tensor becomes three-fold degenerate.
As for the irrelevant perturbation, there seems to be a discrepancy between $\delta x_\sigma$ in Fig.~\ref{Potts_sigma} and Eq.~\eqref{Potts_perturbation}. The data points are scattered for small system sizes and not precisely on the fitting lines. This is due to the leading irrelevant operator we have not considered.
We can identify it as $T_{\text{cyl}}^2+\bar{T}_{\text{cyl}}^2$ as followings.  Just as we did in the left panel of Fig.~\ref{Flow_Ising}, the contributions from $g_X$ can be eliminated by combining $\delta x_\sigma$ and $\delta x_\epsilon$. The OPE coefficients for the 3-state Potts model are known, and the ratio of the two OPE coefficients is ${c_{\epsilon\epsilon X}}/{c_{\sigma\sigma X}}=36$\cite{DOTSENKO1984312,DOTSENKO1985691,DOTSENKO1985291,FUCHS1989303,esterlis2016closure}. Thus, the origin of the "scattering" shall be observed by plotting $36\delta x_\sigma-\delta x_\epsilon$.

\par{}Figure.~\ref{Potts_irrelevant} displays the result for the high-temperature phase. It is now obvious that the scattering of Fig.~\ref{Potts_sigma} comes from the $L^{-2}$ perturbation denoted with the red dotted line. Also, it has a conformal spin $s$ because it flips a sign at each step and $s\equiv4$ (mod 8)~\footnote{For each iteration, the lattice rotates by 45 degrees, and it corresponds to the conformal transformation $w=e^{\frac{i\pi}{4}}z$ on a complex plane.
As the irrelevant perturbations $T_{\text{cyl}}^2$ and $\bar{T}_{\text{cyl}}^2$ have a conformal spin $4$ and $-4$, they get an additional factor $(e^{\frac{i\pi}{4}})^4=(e^{\frac{i\pi}{4}})^{-4}=-1$ for an odd number of steps. We can see this by plotting the data from even steps (original) and odd steps (tilt) separately.}. As a result, we can conclude the irrelevant operator has the conformal weights as $(h,\bar{h})=(4,0)$ and $(0,4)$,
which are $T_{\text{cyl}}^2$ and $\bar{T}_{\text{cyl}}^2$. Finally, the effective Hamiltonian of the critical 3-state Potts model on the square lattice can be constructed as 
\begin{align}
    H=H^*_{\rm Potts}+\int_0^Ldx\left[g_X\Phi_{\frac{7}{5},\frac{7}{5}}(x)+g_T (T_{\text{cyl}}^2+\bar{T}_{\text{cyl}}^2)\right].
\end{align}

\subsection{Finite-Entanglement scaling}
\par{}At the critical temperature of the Ising model, the finite-$D$ effect proves to be a perturbation from the thermal operator. Let us verify it for the critical 3-state Potts model. Due to the irrelevant perturbations from $T_{\text{cyl}}^2+\bar{T}_{\text{cyl}}^2$, the finite-$D$ effects are clearer for $x_\epsilon(L)$ as seen in Fig.~\ref{Potts_sigma}($b$). This is shown in Fig.~\ref{FES_Ising} of the main text. Here, we demonstrate that $\delta x_\sigma(L)$ also shows the universal behavior with $L/\xi(D)$.  
\begin{figure}[tb]
    \centering
    \includegraphics[width=86mm]{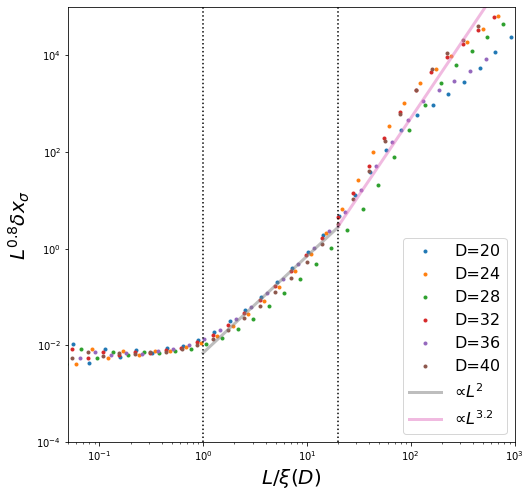}
\caption{Rescaled $\delta x_\sigma$ by $\xi(D)=D^\kappa$ at the critical temperature. The resulting data collapse onto a universal function that is independent of $L/\xi(D)$. If $L/\xi(D) < 1$, the system is in the FSS region, while if $L/\xi(D)\geq 1$, it is in the FES region. In the FES region, the scaling of the first-order and second-order perturbations are indicated by a gray and pink line, respectively. $x_\sigma$ is computed as an average value of the first and second excitation energy.}
    \label{FES_Potts}
\end{figure}
Figure.~\ref{FES_Potts} shows the rescaled correction to $\delta x_\sigma(L)$. For $L>\xi(D)$, the perturbation grows as $L^{2}$ denoted by a gray line, which means that the emergent perturbation scales as $L^{1.2}$.  Compared with Eq.~\eqref{Potts_perturbation}, it is clear that the emergent perturbation is from the thermal operator. However, as the system size increases, the second-order perturbation becomes predominant as shown with a pink line. As $\epsilon$ is the most relevant operator that is permitted by symmetry, it supports our conjecture stated in the main text.
\end{document}